\documentclass[aps,pra,reprint,superscriptaddress,floatfix]{revtex4-2}

\usepackage{amsmath}
\usepackage{amssymb}
\usepackage{amsthm}
\usepackage{bbm}
\usepackage{color}
\usepackage[T1]{fontenc}
\usepackage{graphicx}
\usepackage{hyperref}
\usepackage[utf8]{inputenc}
\usepackage{mathtools}
\usepackage{orcidlink}
\usepackage{physics}
\usepackage{times}
\usepackage{txfonts}
\usepackage{soul}

\hypersetup{
    colorlinks=true,linkcolor=blue,citecolor=blue,
    filecolor=blue,urlcolor=blue,breaklinks=true
}

\theoremstyle{definition}

\def\be{\begin{equation}}
\def\ee{\end{equation}}

\def\bc{\begin{center}}
\def\ec{\end{center}}
\def\bal{\begin{align}}
\def\eal{\end{align}}

\newcommand{\uepg}{
  Departamento de Matemática e Estatística,
  Universidade Estadual de Ponta Grossa,
  84030-900 Ponta Grossa, Paraná, Brazil
}

\newcommand{\qpqi}{
  QPQI Group,
  Universidade Estadual de Ponta Grossa,
  84030-900 Ponta Grossa, Paraná, Brazil
}

\newcommand{\ufpr}{
  Departamento de Física,
  Universidade Federal do Paraná,
  81531-980 Curitiba, Paraná, Brazil
}

\newcommand{\orcidalana}{\orcidlink{0000-0002-9635-0093}}
\newcommand{\orcidthiago}{\orcidlink{0009-0001-1654-0330}}
\newcommand{\orcidfabiano}{\orcidlink{0000-0001-5383-6168}}

\begin{document}

\title{Discrete-time quantum walks with energy-dependent coins}

\author{Alana Spak dos Santos\orcidalana}
\email{alana.spak@ufpr.br}
\affiliation{\ufpr}

\author{Thiago T. Tsutsui\orcidthiago}
\email{takajitsutsui@gmail.com}
\affiliation{\qpqi}

\author{Fabiano M. Andrade\orcidfabiano}
\email{fmandrade@uepg.br}
\affiliation{\qpqi}
\affiliation{\uepg}

\date{\today}

\begin{abstract}
In this work, we extend the scattering quantum walk (SQW) framework to a lattice of energy-dependent point interactions. This yields, within the coined quantum walk (CQW) formalism, a coin operator that is directly related to the scattering matrix of zero-range potentials. The model thus provides a discrete-time quantum-walk (DTQW) analog of a periodic array of point interactions of the Kronig–Penney type, where the walker's wavenumber serves as a continuous, physically transparent control parameter for the coin operation. We analyze the spectra and the dynamics of position probability and entanglement, yielding distinct results for specific energies and point interactions. We relate the spectral structure to the spatial probability distribution and explicitly characterize the long-time entanglement behavior for each point interaction. The transmission modulus determines the quasienergy gap, bandwidth, and maximum group velocity, while also controlling the long-time coin-position entanglement for the initial state considered. The four families of one-dimensional point interactions ($\delta$, $\delta'$, crossed and asymmetric) realize qualitatively distinct transmission profiles and span the full range of behavior, including enhanced or strongly suppressed spreading and oscillatory entanglement.
\end{abstract}


\maketitle

\section{Introduction}

Quantum walks (QWs) represent the quantum versions of the usual
classical random walks and are intimately related to quantum
computation.
In fact, they represent universal models for quantum computation
\cite{Childs2009,Lovett2010,Childs2013}.
In recent decades, this area has produced important achievements in both the theoretical and experimental branches \cite{Schreiber2011,Sansoni2012,Crespi2013,Tao2021,Arute2019}.
QWs are a better option for algorithm development
than classical ones \cite{Shenvi2003,Zhang2022}.
In addition to quantum algorithms, we can also mention the applications
of QWs in quantum cryptography
\cite{Yu-Guang2015,Ahmed2020,El-Latif2021,Yining2022},
topological phases
\cite{Takuya2012,Asboth2013,PANAHIYAN2020,Yali2021,Panahiyan2021},
quantum phase transitions \cite{wang2019},
and secure quantum memory \cite{Chandrashekar_2015}.

The theoretical implementation of QWs makes use of graphs, mathematical
objects consisting of sets of vertices and edges.
QWs can be discussed in a continuous- or discrete-time framework \cite{Kempe2003}.
In continuous-time QWs (CTQWs), the quantum dynamics uses the adjacency
matrix of the underlying graph as the Hamiltonian of the system
\cite{Farhi1998}.
Discrete-time QWs (DTQWs) can be formulated in two distinct but
equivalent frameworks \cite{PRA.80.052301.2009} depending on the
dynamical process emphasized in the graph:
(i) the purely stochastic one associated with the choice of direction
at the vertices of the graph; or
(ii) the purely deterministic one associated with the ballistic motion
at the edges of the graph.
The first discrete-time model, associated with scenario (i),
is the coined quantum walk (CQW)
\cite{Inproceedings.2001.Watrous,Inproceedings.2001.Ambainis}, in which the quantum states of the system are defined on the vertices of the graph.
The second discrete-time model, associated with scenario (ii),
is the scattering quantum walk (SQW)
\cite{Conference.2005.Feldman,Feldman2004,Hillery2003}, in which the quantum states are defined at the edges of the graph.
In what follows, we focus on discrete-time QW models.

The connection between SQW and quantum graphs \cite{Book.2012.Berkolaiko,PR.647.1.2016} provides a natural framework in which the local scattering amplitudes can depend on the wavenumber. In particular, the Green's function approach to SQWs established a connection between SQWs and generalized Kronig-Penney systems \cite{PRSLA.130.499.1931,PRL.74.3503.1995} and naturally accommodates energy-dependent scattering amplitudes \cite{PRA.84.042343.2011}.
The unitary equivalence between the coined and scattering formulations was also developed for arbitrary position-dependent transition amplitudes and illustrated for several lattice topologies \cite{JPAMT.46.165302.2013}.
Related connections between continuous scattering problems and DTQWs have been investigated from a transfer-matrix and semiclassical perspective \cite{Higuchi2021}.
Thus, the use of point interactions to generate an energy-dependent DTQW is naturally rooted in the scattering formulation rather than being an independent choice of coin parameters.

In this work, we build on this established scattering framework and focus on consequences not addressed in those constructions, namely the spectral and quantum-information structure of the resulting homogeneous energy-dependent walk. For a periodic one-dimensional array of point interactions, we analyze the quasienergy spectrum, spatial spreading, and coin-position entanglement, and derive explicit relations between the transmission modulus and the minimum quasienergy gap, bandwidth, and maximum group velocity. We further show that, for the initial state considered, the long-time coin-position entanglement is controlled by the scattering amplitudes. In contrast, the transient oscillations of the entanglement carry information about the quasienergy gap. The broader scattering-walk framework has also been developed for general scattering matrices and other lattice geometries \cite{APN.396.517.2018,PRB.108.94303.2023}. Here we specialize to one-dimensional point interactions to expose these spectral and entanglement relations analytically. 

The paper is organized as follows.
In Sec. \ref{sec:models}, we briefly review the CQW and SQW models and
their unitary equivalence.
In Sec. \ref{sec:edcoins}, we formulate the energy-dependent DTQW and analyze the quasienergy spectra and the spatial distribution.
In Sec. \ref{sec:entanglement}, we discuss the entanglement, including its asymptotic behavior and spectral information for the four potentials considered.
Finally, in Sec. \ref{sec:conclusion}, we present our conclusions.

\section{Discrete-time quantum walk models}
\label{sec:models}

In this section, we briefly review the CQW and SQW models.
Although these models are formulated in different Hilbert spaces and
admit distinct physical interpretations, they are unitarily equivalent.
This equivalence is central to our work, as it allows us to exploit the
physical intuition and generality of the SQW model while also relating
the results to the well-known CQW model.
Additionally, this mapping allows us to see the bipartite structure of
the SQW more clearly, which, in part, makes the entanglement computation
direct \cite{Carneiro2005}.
The description of these models makes use of graphs to represent the
system.

\subsection{Coined quantum walks}
\label{subsec:CQW}

In the QW model, in which the purely stochastic process is taken as primary, the Hilbert space vectors that describe the state of the system
are given by the orthonormal basis states $\{\ket{j},\, j \in
\mathbb{Z}\}$.
These quantum states are defined on the vertices of the graph and span
the position Hilbert space $\mathcal{H}_p= L^{2}(\mathbb{Z})$.
Additionally, for each $j$, there is a quantum coin space that possesses two orthonormal states $\ket{+}=\left(1,0\right)^T$ and $\ket{-}=\left(0,1\right)^T$, corresponding to right and left, respectively, and spans the two-dimensional coin space $\mathcal{H}_c$.
Therefore, the Hilbert space for the entire system is $\mathcal{H}_{\rm cqw}=\mathcal{H}_p \otimes \mathcal{H}_c = L^{2}(\mathbb{Z})\otimes L^2(\mathbb{Z}_2)$, and the orthonormal basis states are  described by $\ket{j}\otimes\ket{\sigma}$, with $ j \in \mathbb{Z}$ and $\sigma =\pm$.
A single time step in the CQW consists of the application of two unitary operators: the application of a rotation in the coin space by means of a unitary coin operator $\mathbf{C}_c \in U(2)$, with $U(2)$  as the set of $2\times 2$ unitary matrices, followed by a unitary conditional shift operator,
\begin{equation} \label{eq:trans}
  \mathbf{S}_p=\sum_{j\in \mathbb{Z}}
  \left(
  \dyad{j+1}{j}\otimes\dyad{+}
  +
  \dyad{j-1}{j}\otimes\dyad{-}
  \right).
\end{equation}
Thus, a time step in CQW is given by the unitary operator
\begin{equation}
  \mathbf{U}_c=
  \mathbf{S}_{p}(\mathbf{\mathbbm{1}}_{p}\otimes \mathbf{C}_{c}).
\end{equation}
After $n$ time steps, the state of the system is given by
\begin{equation}
  \ket{\phi(n)} = \mathbf{U}_c^{n}\ket{\phi(0)},
\end{equation}
with $\ket{\phi(0)}$ an initial quantum state.

A frequently used coin operator is the Hadamard coin
\begin{equation}
  \mathbf{H}_{c} =
  \frac{1}{\sqrt{2}}
  \begin{pmatrix*}[r]
    1 & 1\\
    1 & -1
  \end{pmatrix*},
\end{equation}
which generates an equal superposition of states $\ket{+}$ and $\ket{-}$,
\begin{equation}
  \mathbf{H}_{c} \ket{\pm}=\frac{1}{\sqrt{2}}(\ket{+}\pm\ket{-}).
\end{equation}
When the Hadamard coin is used as the coin operator, the resulting QW is called the Hadamard quantum walk.

\subsection{Scattering quantum walks}
\label{subsec:SQW}

The QW model in which the deterministic process is taken as the primary one was proposed in \cite{Hillery2003,Feldman2004} and is based on an interferometric analogy.
The vertices are treated as optical elements with $2D$ ports, where $D$ is the degree of the vertices, and the edges correspond to the paths a photon can follow through the interferometer.
The states are defined on the edges of the graph, and the vertices are beam splitters, so the dynamics takes place on the edges, with the vertices as scattering centers.

On the line, the orthonormal basis states are $\ket{\sigma,j}$, with $\sigma=\pm$  representing the propagation direction of the quantum walker, $+$ ($-$) representing the right (left) propagation direction, and $j\in\mathbb{Z}$ as the vertex.
Then, the Hilbert space is $\mathcal{H}_{\rm sqw}=L^2(\mathbb{Z}\times\mathbb{Z}_2)$.
The translation operation is given by the application of a unitary operator $\mathbf{U}_s$ and depends on the direction of the state
\begin{align}
  \label{eq:ev}
  \mathbf{U}_s\ket{+,j} = {}
  &
    r\ket{-,j-1} +
    t\ket{+,j+1}, \nonumber \\
  \mathbf{U}_s\ket{-,j} = {}
  &
    -r^{*}\ket{+,j+1}
    +t^{*}\ket{-,j-1},
\end{align}
where $r$ and $t$ are the reflection and transmission
amplitudes that satisfy unitary conditions.
Thus, the state after $n$ time steps is obtained by successive
applications of $\mathbf{U}_s$ on the initial state $\ket{\psi(0)}$,
\begin{equation}
  \ket{\psi(n)} = \mathbf{U}_{s}^n\ket{\psi(0)}.
\end{equation}
The Hadamard quantum walk can be obtained from the SQW using $r = t = 1/\sqrt{2}$ \cite{Hillery2003}.

In contrast to the CQW, the SQW model does not explicitly exhibit a
tensor-product decomposition into position and internal degrees of
freedom.
Nevertheless, as shown below, this apparent structural difference does
not imply distinct physical dynamics.

\subsection{Unitary equivalence between CQW and SQW}

When the SQW was first introduced in \cite{Feldman2004}, the authors established its unitary equivalence with the CQW on the line.
The equivalence was later generalized to arbitrary topologies (graphs) in \cite{PRA.80.052301.2009}, proving that it is always possible to map one
model onto the other in any topology.
Thus, depending on the specific properties we wish to investigate, we may work with either the SQW or the CQW model, or even a mixed approach.

Let us start by reminding that the Hilbert space of the CQW is $\mathcal{H}_{\rm cqw}= L^{2}(\mathbb{Z})\otimes L^{2}(\mathbb{Z}_2)$, which is identical to the Hilbert space of the SQW, $\mathcal{H}_{\rm sqw}= L^{2}(\mathbb{Z}\times\mathbb{Z}_2)$.
The unitary mapping between the SQW and the CQW is established by the isomorphic operator $\mathbf{E}: \mathcal{H}_{\rm sqw} \to \mathcal{H}_{\rm cqw}$, given by
\begin{equation}
\mathbf{E}=\sum_{j \in \mathbb{Z}}
\left(
\ket{j}\otimes\ketbra{+}{+,j}+
\ket{j}\otimes\ketbra{-}{-,j}
\right),
\end{equation}
in such a way that
\begin{equation}
  \label{eq:E_operator}
  \mathbf{E} \ket{\pm,j} =  \ket{j}\otimes\ket{\pm},
\end{equation}
which effectively translates as the incoming scattering state ($\pm$)
at a given position $j$ being mapped to the outgoing coin state
($\pm$) at the same position.
In this framework, the time evolution operators of the two models are
related by the transformation
$\mathbf{U}_c=\mathbf{E}\mathbf{U}_s\mathbf{E}^{-1}$.
Thus, the two QW models are related by a unitary transformation and are
dynamically equivalent.

Any apparent difference between the two formulations arises from the
operational definition of probabilities.
In the CQW, the probability of finding the walker at vertex $j$ after
$n$ steps is obtained by summing over the squared amplitudes for the
coin states  $\ket{j}\otimes\ket{-}$ and $\ket{j}\otimes\ket{+}$,
\begin{equation}
P_j(n) =
|(\bra{j}\otimes\bra{-}) \ket{\phi(n)}|^2
+
|(\bra{j}\otimes\bra{+})\ket{\phi(n)}|^2.
\end{equation}
In the SQW, by the inverse mapping $\mathbf{E}^{-1}$, these states correspond to states on different edges, $\ket{-,j}$ and $\ket{+,j}$,
respectively.
Thus, the probability in the edge $(j-1,j)$ is given by the addition of the square amplitudes of states $\ket{+,j}$ and $\ket{-,j-1}$,
\begin{equation}
\label{eq:pedge}
P_{(j-1,j)}(n) =  |\braket{-,j-1}{\psi(n)}|^2+|\braket{+,j}{\psi(n)}|^2
\end{equation}
Moreover, when the SQW state is properly projected onto the vertices, the resulting probability distribution coincides exactly with that of the
CQW, so that the probability $P_j(n)$ can also be obtained by
\begin{equation}
\label{eq:pvertex}
  P_j(n) =
  |\bra{-,j}\ket{\psi(n)}|^2
  +
  |\bra{+,j}\ket{\psi(n)}|^2.
\end{equation}

\begin{figure}[t]
  \centering
   \includegraphics[width=\columnwidth]{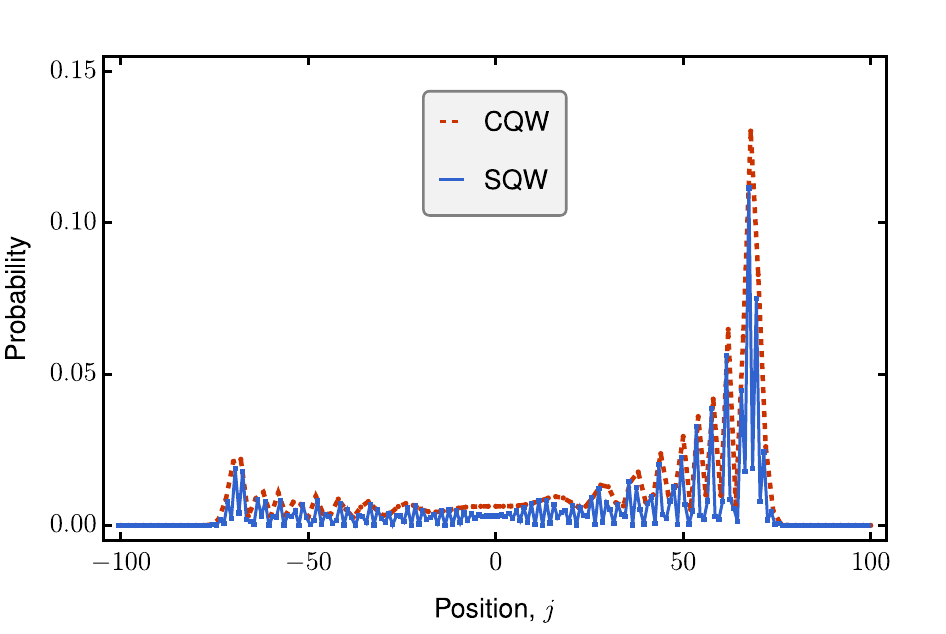}
   \caption{%
    The blue curve is obtained from Eq. \eqref{eq:pedge} and shows the probabilities for the SQW with $n=100$ steps, $\ket{\psi(0)}=\ket{+,0}$ and $r=t=1/\sqrt{2}$.
    The dashed red curve is obtained from Eq. \eqref{eq:pvertex} and shows the probabilities for the CQW obtained from the projection of the state $\ket{\psi(n)}$ obtained from the SQW onto the vertices (blue curve).
  }
  \label{fig:fig1}
\end{figure}

In Fig. \ref{fig:fig1}, we show the comparison between the
probabilities for SQW and CQW obtained from the evolution of SQW, considering $\ket{\psi(0)}=\ket{+,0}$ as the initial state and using $r = t = 1/\sqrt{2}$.
Therefore, the distinction between the two models is operational rather than dynamical.
Although the CQW formulation is particularly convenient for defining bipartite structures and entanglement measures, the SQW formulation provides a more direct physical interpretation in terms of wave propagation and scattering processes.

Thus, as stated above, this equivalence allows us to adopt a mixed approach: we use the SQW framework, which is especially suitable for introducing energy-dependent scattering amplitudes, while consistently analyzing entanglement properties through the CQW representation.

\section{Quantum walks with energy-dependent coins}
\label{sec:edcoins}

As discussed above, the analogy between QWs and interferometric networks allows us to formulate QWs within the framework of scattering theory.
However, despite its broad applicability, the standard SQW formalism does not address the most general scattering scenario, as it assumes energy-independent scattering amplitudes.
To overcome this limitation, we reinterpret the interferometric network in the spirit of \emph{quantum graphs} \cite{Book.2012.Berkolaiko,PR.647.1.2016}, in which the vertices are zero-range potentials, and the edges correspond to free propagation regions.
Within this framework, the quantum walker carries a well-defined energy, naturally leading to energy-dependent scattering.
This viewpoint is consistent with the Green's function approach for SQWs described in  \cite{PRA.84.042343.2011}.

\subsection{1D point interactions}
A mathematically rigorous treatment of short-range potentials on the
line is achieved by modeling them as point interactions through the
framework of self-adjoint extensions \cite{Book.2004.Albeverio}.
Within this formalism, solving the free Schrödinger equation
$-d^2\psi(x)/dx^2 = k^2\psi(x)$, with $k=\sqrt{2E}$ ($m=\hbar=1$), for a
single-point interaction at the vertex $j$ (placed at $x=0$ without loss of generality) reduces to imposing the following boundary condition
\cite{JPA.26.427.1993,JPA.30.3937.1997} ($\psi'=d\psi/dx$):
\begin{equation}
  \Psi(0^{+})=\Gamma \Psi(0^{-}),
  \label{eq:bc}
\end{equation}
where
\begin{equation}
\label{eq:Psi}
  \Psi(x)=
  \left(
    \begin{array}{c}
      \psi(x)\\
      \psi'(x)
    \end{array}
  \right),
\end{equation}
with $\Psi(0^{\pm})=\lim_{x \to 0^{\pm}}\Psi(x)$ and
\begin{equation}
  \Gamma=
  \omega
  \left(
    \begin{array}{cc}
      a & b\\
      c & d
    \end{array}
  \right),
\end{equation}
where
\begin{equation}
\label{eq:restrictions}
|\omega|=1
\quad \mbox{and} \quad
a d  - b c = 1, \quad \mbox{with} \quad
a, b, c, d \in\mathbbm{R}.
\end{equation}
From a physical standpoint, the conditions above ensure the conservation
of the probability flux as shown in Appendix \ref{sec:appendix}.

Alternatively, the boundary condition in Eq. \eqref{eq:bc} can be
characterized in terms of the amplitudes of scattering.
Consider a plane wave with wavenumber $k$ incident from the left $(+)$ or the right $(-)$ upon the point interaction at the vertex $j$.
The corresponding scattering wave functions, which satisfy
$-d^2\psi(x)/dx^2 = k^2\psi(x)$ for $x \neq 0$, are given by
\begin{equation}
 \psi^{(\pm)}(x)=\frac{1}{\sqrt{2\pi}}
   \begin{cases}
     e^{\pm ikx}+r^{(\pm)}(k)e^{\mp ikx}, & x \lessgtr 0\\
     t^{(\pm)}(k) e^{\pm ikx}, &  x \gtrless  0.
   \end{cases}
 \label{eq:psi-scattering}
\end{equation}
Inserting Eq. \eqref{eq:psi-scattering} into Eq. \eqref{eq:bc} yields the following general expression for the reflection and transmission
amplitudes:
\begin{subequations}
    \begin{equation}
  r^{(\pm)}(k) = \frac{c \pm ik(d-a) + bk^2} {-c + i k(d+a) + b k^2},
  \label{eq:r-sqw}
\end{equation}
\begin{equation}
  t^{(\pm)}(k) = \frac{2 i k \omega^{\pm 1}}{-c + i k(d+a) + b k^2 },
  \label{eq:t-sqw}
\end{equation}
\end{subequations}
where the reflection and transmission amplitudes are now energy-dependent
through the wavenumber $k$.
Writing $\vartheta(k)= -c +ik (d+a)+bk^2$ for the common denominator and  using $ad-bc=1$, we can write $|t^{(\pm)}(k)| = \sqrt{4k^2/|\vartheta(k)|^2}$ for every point interaction.
Since $|\omega|=1$, we have $|t^{(+)}|=|t^{(-)}|$, and likewise $|r^{(+)}|=|r^{(-)}|$.
So, whenever modulus matters, we can simply write $|t(k)|$ and $|r(k)|=\sqrt{1-|t(k)|^2}$.

\begin{table}[t]
  \caption{\label{tab:table1}
    Point interactions that are analyzed in this work.}
\begin{ruledtabular}
\begin{tabular}{cccccr}
Point interaction & $a$ & $b$ & $c$ & $d$ & $\omega$\\
\hline
delta ($\delta$)  & $1$ & $0$ & $2\gamma$ & $1$ & $1$\\
delta-prime ($\delta'$) & $1$ & $2\gamma$ & $0$ & $1$ & $1$\\
crossed ($\chi$)    & $0$ & $\gamma$ & $-\gamma^{-1}$ & $0$ & $1$\\
asymmetric ($\alpha$)  & $\gamma^{-1}$ & $0$ & $\gamma^{-1}$ & $\gamma$ & $-i$\\
\end{tabular}
\end{ruledtabular}
\end{table}

The scattering amplitudes can be conveniently grouped into the scattering matrix, or $S$-matrix, which maps the \emph{incoming} waves to the \emph{outgoing} waves \cite{Book.1998.Merzbacher}.
For a single-point interaction, the $S$-matrix is thus given by
\begin{equation}
S(k) =
\left(
\begin{matrix}
r^{(+)}(k) & t^{(-)}(k) \\
t^{(+)}(k) & r^{(-)}(k)
\end{matrix}
\right).
\label{eq:S-matrix}
\end{equation}
Conservation of the probability flux implies that $S(k)$ is a unitary matrix, $S(k) \in U(2)$, leading to the relations
\begin{gather}
|r^{(\pm)}(k)|^2 + |t^{(\pm)}(k)|^2 = 1,
\label{eq:rt-relations-qrw}
\\
r^{(+)}(k)^*\, t^{(-)}(k)
+ r^{(-)}(k)\, t^{(+)}(k)^* = 0.
\nonumber
\end{gather}
and the existence of an inverse scattering problem \cite{Book.1989.Chadan} leads to
\begin{equation}
\label{eq:rt-symmetry}
r^{(\pm)}(k)^* = r^{(\pm)}(-k),
\quad \text{and} \quad
t^{(\pm)}(k)^* = t^{(\mp)}(-k).
\end{equation}
Additionally, if we require time-reversal invariance, we must impose
$t^{(+)}(k)=t^{(-)}(k)$, leading to $\omega=\pm 1$.
Unless otherwise stated, we do not impose this latter requirement in the present work.

\begin{figure}[t]
  \centering
  \includegraphics[width=\columnwidth]{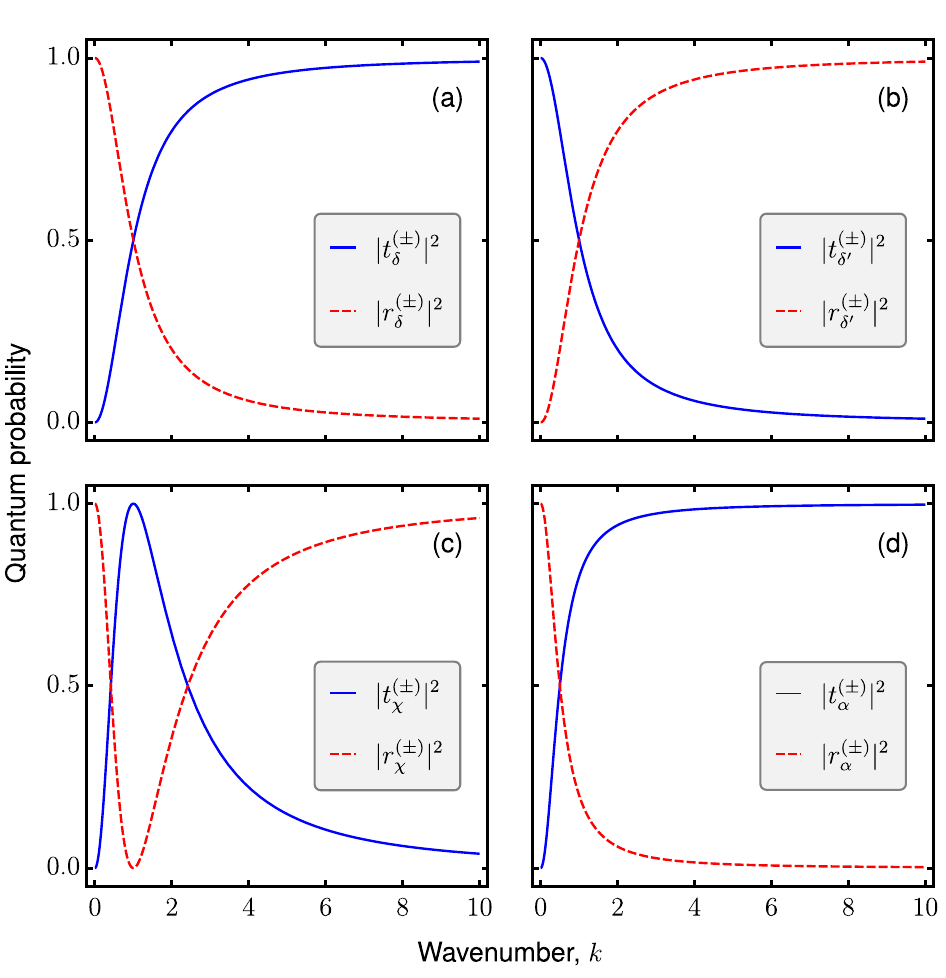}
  \caption{
    Reflection (dashed red line) and transmission (solid blue line) probabilities for $\gamma=1$ as a function of the wavenumber $k$ for the point interactions:
    (a) $\delta$, (b) $\delta'$, (c) $\chi$, and (d) $\alpha$.
    }
  \label{fig:fig2}
\end{figure}

General 1D point interactions can be classified by the behavior of wave function scattering properties \cite{JPA.39.2493.2006}.
Thus, we use this classification and focus on four possible cases:
delta ($\delta$), delta-prime ($\delta'$), crossed ($\chi$), and asymmetric ($\alpha$).
Each case is determined by the set of parameters $a$, $b$, $c$, $d$, and $\omega$, and is summarized in Table \ref{tab:table1}.
In terms of the boundary conditions, the $\delta$ interaction corresponds to the case where the wave function is continuous at the point supporting the interaction  $\psi(0^{+})=\psi(0^{-}):=\psi$, and the first derivative is discontinuous $\psi'(0^{+})- \psi '(0^{-})=2\gamma\psi$.
In the $\delta'$ interaction, the roles of the wave functions and the derivatives are reversed, i.e., the derivative is continuous $\psi'(0^{+})=\psi'(0^{-}):=\psi'$, and the wave function is discontinuous $\psi(0^{+})- \psi(0^{-})=2\gamma\psi'$.
The $\chi$ interaction corresponds to the boundary condition where the wave function behaves as $\psi(0^{+}) = \gamma \psi{'}(0^{-})$ and $\psi'(0^{+})=-\gamma^{-1}\psi(0^{-})$, which means that the value of $\psi$ on one side depends only on the derivative on the other side of the point interaction.
Finally, in the $\alpha$ interaction \cite{PRA.69.052708.2004}, we find that the quantum amplitudes on the right and left are different, which corresponds to a break in time-reversal invariance.
However, although the $\alpha$ interaction breaks time-reversal invariance at the amplitude level, the left- and right-incidence transmission probabilities remain identical.
In Fig. \ref{fig:fig2}, we show the behavior of the reflection (dashed line) and transmission (solid line) coefficients for these four cases as a function of wavenumber $k$.

\subsection{Energy-dependent DTQWs}

\begin{figure}[t]
  \centering
  \includegraphics[width=\columnwidth]{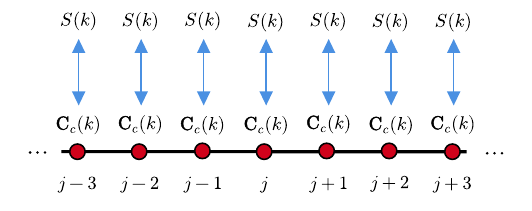}
  \caption{
    Schematic representation of the point interactions in a linear
    quantum graph, and the associated $S$-matrix and energy-dependent coin
    operators.
  }
  \label{fig:fig3}
\end{figure}

\begin{figure*}
  \centering
  \includegraphics[width=\textwidth]{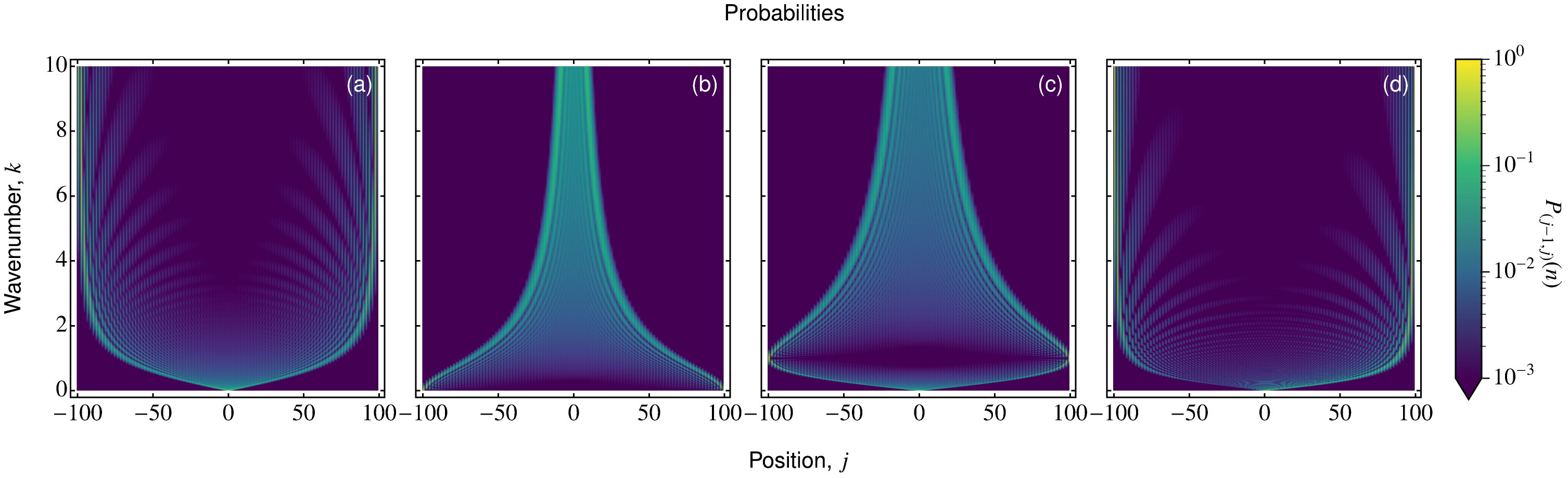}
  \caption{
    Density plot of the probability distribution for $k\in[0,10]$,
    $\gamma=1$ and initial state
    $\ket{\psi(0)}=1/\sqrt{2}(\ket{-,0}+\ket{+,0})$.
    We consider the (a) $\delta$, (b) $\delta'$, (c) $\chi$,
    and (d) $\alpha$ interactions.}
  \label{fig:fig4}
\end{figure*}

Thus, in the most general scattering scenario, the scattering amplitudes depend on energy.
Based on this, an energy-dependent SQW is defined by the energy-dependent one-step translation operator
\begin{equation}
  \label{eq:ev-gen}
  \mathbf{U}_s(k)\ket{\pm,j} =
    r^{(\pm)}(k)\ket{\mp,j \mp 1} +
    t^{(\pm)}(k)\ket{\pm,j \pm 1},
\end{equation}
with $r^{(\pm)}(k)$ and $t^{(\pm)}(k)$ satisfying
Eqs. \eqref{eq:rt-relations-qrw} and \eqref{eq:rt-symmetry}.
Moreover, from the mapping in Eq. \eqref{eq:E_operator}, it is straightforward to obtain the energy-dependent coin operator
\begin{equation}
    \mathbf{C}_c(k) \ket{\pm} =
    r^{(\pm)}(k) \ket{\mp} + t^{(\pm)} (k) \ket{\pm}
\end{equation}
or, in its matrix form,
\begin{equation}
  \mathbf{C}_{c}(k)=
  \left(
    \begin{array}{cc}
     t^{(+)}(k) & r^{(-)}(k) \\
     r^{(+)}(k) & t^{(-)}(k)
    \end{array}
  \right).
\label{eq:Coin_scatt_1D}
\end{equation}
Then, we can express the relationship between the coin operator and the $S$-matrix as follows
\begin{equation}
\mathbf{C}_c(k) =  \sigma_x S(k),
\end{equation}
with $\sigma_x$ the Pauli $x$ matrix.
Since $\sigma_x$ is a unitary matrix, the resulting coin operator is, as required, unitary as well.
In this sense, mapping the energy-dependent SQW to the CQW leads to the energy-dependent coin operator, whose structure is directly determined by the scattering properties of the underlying point interaction.
Therefore, unlike conventional CQWs, where the coin operator is introduced as an abstract $U(2)$ operator, the present formulation derives the coin directly from a physically motivated scattering process, leading to an energy-dependent unitary operator
$\mathbf{U}_c(k)= \mathbf{S}_{p}[\mathbf{\mathbbm{1}}_{p}\otimes \mathbf{C}_{c}(k)]$.

The resulting dynamics thus suggests a connection to quantum scattering and transport in generalized Kronig-Penney models, which consist of a periodic array of point interactions \cite{PRSLA.130.499.1931}.
In Fig. \ref{fig:fig3}, we schematically represent a linear quantum graph with point interactions, the associated scattering matrix at each vertex, and the corresponding energy-dependent coin operator.
It should be stressed that at a fixed energy, the scattering matrix $S(k)$ and the corresponding coin $\mathbf{C}_c(k)$ are the same at every vertex, so the walk remains a homogeneous DTQW with a particular $U(2)$ coin.
The wavenumber $k$ enters as a parameter, not as a dynamical variable, and the scattering problem constrains the coin to a specific one-parameter $U(2)$ realization associated with each point interaction of Table \ref{tab:table1}.

\subsection{Energy-dependent DTQW dynamics}

To illustrate how the scattering matrix determines the DTQW evolution as a function of energy, we discuss the four cases in Table \ref{tab:table1}.
Figure \ref{fig:fig4} shows the probability distribution results.
In the $\delta$ case [Fig. \ref{fig:fig4}(a)], as observed in Fig. \ref{fig:fig2}(a), the transmission coefficient increases as $k$ increases, so the quantum walker exhibits increasingly pronounced ballistic spreading.
An increase in the $\gamma$ intensity leads to a decrease in the transmission coefficient for a given $k$, resulting in slower spreading of the quantum walker.
In the case of a $\delta'$ [Fig. \ref{fig:fig4}(b)], the transmission coefficient decreases as $k$ increases [see Fig. \ref{fig:fig2}(b)], causing the spatial spreading to become increasingly suppressed.
Increasing $\gamma$, we observe that the walker becomes even more
localized around the initial position.
The $\chi$ case [Fig. \ref{fig:fig4}(c)] is interesting as the transmission coefficient displays a maximum at $k=k_{max}=\sqrt{|1-ad|}/|b|=1/\gamma$ \cite{JPA.39.2493.2006}.
For the parameter values considered, the transmission coefficient peaks at $k_{max}=1$ [Fig. \ref{fig:fig2}(c)], and the walker exhibits ballistic spreading [see Fig. \ref{fig:fig4}(c)].
For larger values of $k$, the transmission coefficient decreases, and the walker tends to localize near the initial position.
Increasing $\gamma$, $k_{max}$ decreases, and the transmission coefficient diminishes for the same value of $k$. In this way, the walker behaves ballistically for lower values of $k$.
The $\alpha$ case [Fig. \ref{fig:fig2}(d)] has behavior similar to the $\delta$ case [see Fig. \ref{fig:fig4}(d)], with the difference that the walker spreads out more rapidly because the transmission coefficient is greater for the same $k$ when compared to the delta case.
In this case, as $\gamma$ increases, the transmission coefficient has an asymptotic value of $|t_{k\to\infty}|^2=4/(d+d^{-1})^2=4\gamma^2/(1+\gamma^2)^2$ \cite{JPA.39.2493.2006}, which equals unity for $\gamma=1$ (the value used in Fig. \ref{fig:fig2}), causing the walker to be located at a fixed distance, even with the increase of $k$.

\subsection{Spectral properties of the energy-dependent DTQW}

The time-independent aspect of the energy-dependent unitary evolution operator $\mathbf{U}_c(k)$ in the CQW allows us to identify it as a Floquet operator \cite{Romanelli2005,Vakulchyk2017,Buarque2019}.
In this context, for a fixed $k$, we associate $\mathbf{U}_c(k)$ with an effective Hamiltonian $H_{\text{eff}}(k)$, such that $\mathbf{U}_c(k) = e^{-i H_{\text{eff}}(k)}$.
Consequently, we obtain an energy-dependent Floquet quasienergy spectrum \cite{Vakulchyk2017,LoGullo2017,Buarque2019,Gong2022}
\begin{equation}
    \epsilon_i(k)=i\log\left[\mu_i(k)\right] \pmod{2\pi},
\end{equation}
where $\mu_i(k)$ and $\epsilon_i(k)$ are the eigenvalues of $\mathbf{U}_c(k)$ and $H_\text{eff}(k)$, respectively. This analysis enables the investigation of a quasienergy spectrum that depends on $k$ for all four types of point interactions.
Additionally, the structure of the spectra and the walker's group velocity allow a direct connection to the transmission properties of the scattering.

For a general time-evolution operator, this study requires numerical diagonalization.
However, given spatial homogeneity, we can follow the approach in Ref. \cite{LoGullo2017} to obtain analytical expressions for the quasienergy eigenvalues.
The method consists of introducing the quasimomentum space, with base states $\ket{\kappa}$ satisfying $\int \frac{d \kappa}{2\pi} \ketbra{\kappa}= \mathbbm{1}$.
Consequently, employing Eq. \eqref{eq:Coin_scatt_1D}, we obtain
\begin{align}
    \mathbf{U}_c(k)= {}
    &
    \int \frac{d \kappa}{2\pi} \ketbra{\kappa}\otimes\left(e^{-i\kappa}\ketbra{-} +e^{i\kappa}\ketbra{+}\right) \mathbf{C}_c(k) \nonumber \\
    = {}
    &
    \int \frac{d \kappa}{2\pi} \ketbra{\kappa}\otimes
    \left(\begin{array}{cc}
    e^{i\kappa}t^{(+)}(k) & e^{i\kappa}r^{(-)}(k)\\
    e^{-i\kappa}r^{(+)}(k) & e^{-i\kappa}t^{(-)}(k)
\end{array}\right). \label{eq:kappa_matrix}
\end{align}
Thus, the problem reduces to diagonalizing the $2\times2$ matrix in Eq. \eqref{eq:kappa_matrix} over the interval $\kappa\in[-\pi,\pi]$.
In this context, we obtain the secular equation
\begin{align} \mu^{2}(\kappa,k)-&\mu(\kappa,k)\left[e^{i\kappa}t^{(+)}(k)+e^{-i\kappa}t^{(-)}(k)\right]
   + D(k)=0,
\end{align}
where $D(k)=\det[\mathbf{C}_c(k)]= t^{(+)}(k)t^{(-)}(k)-r^{(-)}(k)r^{(+)}(k)$.
Since the matrix is unitary, we may write $D(k)=e^{i\phi(k)}$, and the Floquet quasienergies are given by
\begin{equation} \label{eq:quasi}
   \epsilon_{\pm}(\kappa,k) =-\frac{\pi}{2}+\arg\vartheta(k)\pm \Omega(\kappa,k),
\end{equation}
with
\begin{equation}
    \Omega(\kappa,k) =\arccos\left\{\left|t(k)\right|\cos\left[\kappa+\arg\left(\omega\right)\right]\right\}
\end{equation}
The double in the quasienergy corresponds to the two quasienergy bands, which are associated with the internal coin degrees of freedom \cite{LoGullo2017}, and satisfy
\begin{equation}
  \epsilon_{+}(\kappa,k)+\epsilon_{-}(\kappa,k)=-\phi(k), \pmod{2\pi},
\end{equation}
and
\begin{equation}
\epsilon_{+}(\kappa,k)-\epsilon_{-}(\kappa,k)=2\Omega(\kappa,k), \pmod{2\pi}.
\end{equation}

\begin{figure}[b]
  \centering
  \includegraphics[width=\columnwidth]{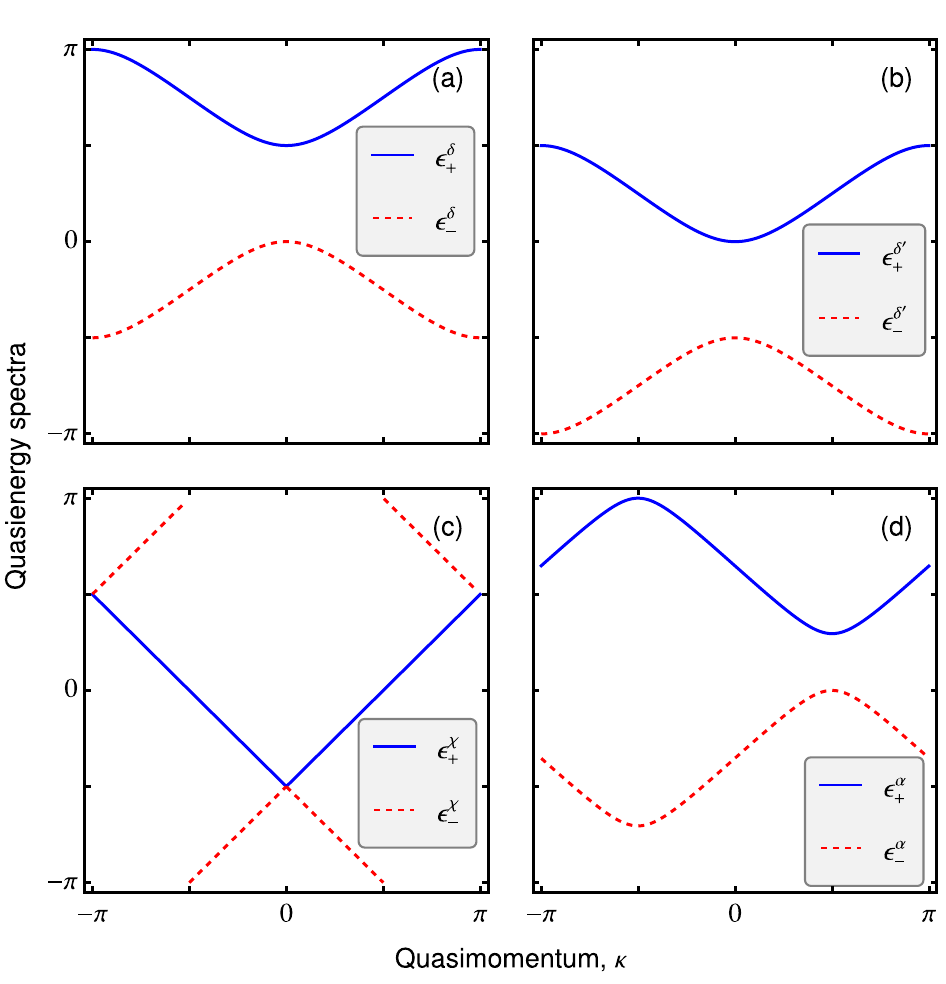}
  \caption{
   Quasienergy eigenvalues for the point interactions with $k=\gamma=1$:
    (a) $\delta$, (b) $\delta'$, (c) $\chi$, and (d) $\alpha$.
    }
  \label{fig:fig5}
\end{figure}

\begin{figure}[t]
  \centering
  \includegraphics[width=\columnwidth]{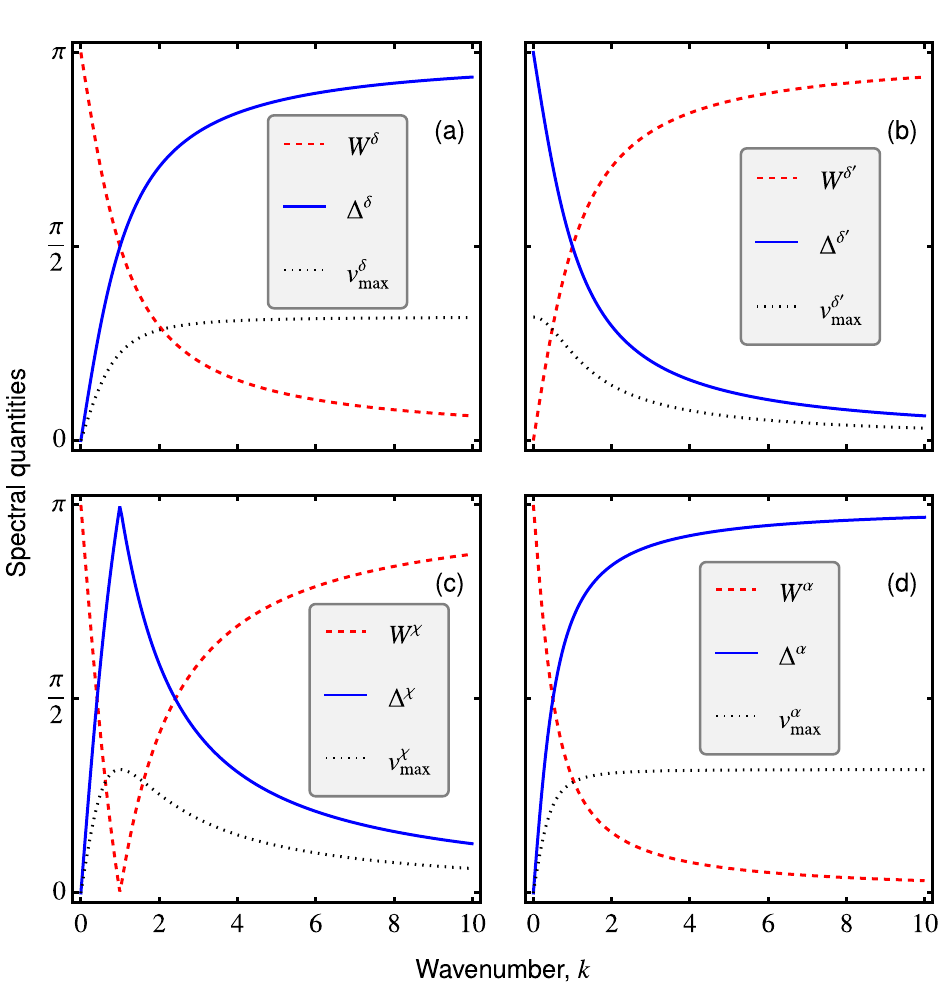}
  \caption{
   Minimum band gap, bandwidth, and maximum group velocity for the point interactions with $\gamma=1$:
    (a) $\delta$, (b) $\delta'$, (c) $\chi$, and (d) $\alpha$.
    }
  \label{fig:fig6}
\end{figure}

\begin{figure*}
  \centering
  \includegraphics[width=\textwidth]{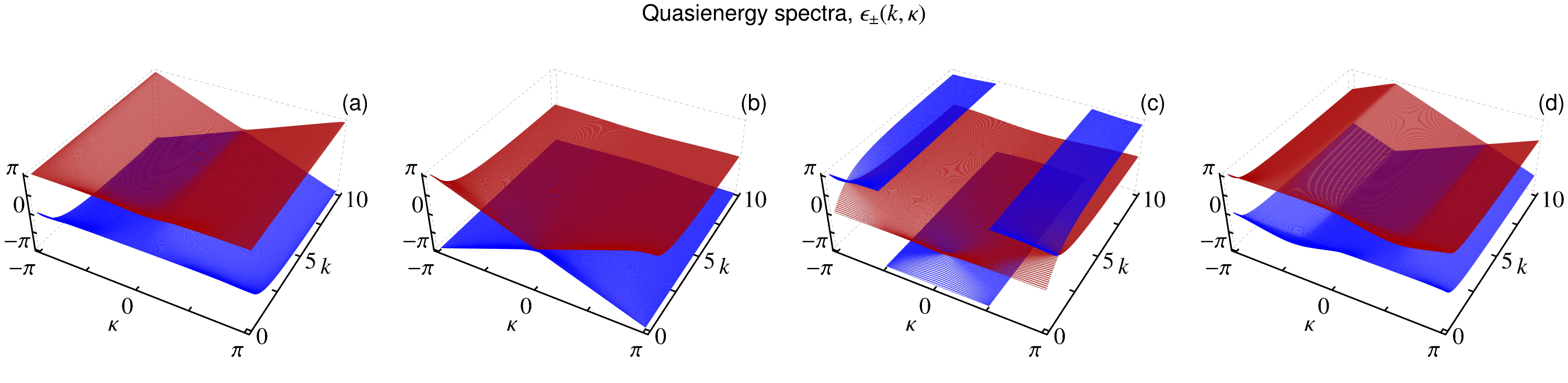}
  \caption{
    3D plot of the energy spectra for $k\in[0,10]$, $\gamma=1$ and initial state     $\ket{\psi(0)}=(\ket{-,0}+\ket{+,0})/\sqrt{2}$.
    We consider the four cases discussed in the text:
    (a) $\delta$, (b) $\delta^{'}$, (c) $\chi$, and (d) $\alpha$.}
  \label{fig:fig7}
\end{figure*}

We note the dependence on the transmission magnitude,
$|t(k)|=\sqrt{{4k^2}/{|\vartheta(k)|^2}}$, and two useful quantities can
be derived  from Eq. \eqref{eq:quasi}: the minimum band
gap and the bandwidth, which are given, respectively, by
\begin{align}
  W(k) = {}
  &
    \min_{\kappa}\epsilon_{\text{gap}}(\kappa,k) \nonumber \\
  =  {}
  &
    2\arccos\left[|t(k)|\right],
\end{align}
and
\begin{align}
  \Delta(k) = {}
  &
    \epsilon_{\text{width}}(k) \nonumber \\
    = {}
  &
    \pi - 2\arccos\left[|t(k)| \right].
\end{align}
They relate the structure of the quasienergy spectrum to the transport properties of the energy-dependent CQW and satisfy the complementarity relation
$W(k)+\Delta(k) = \pi$.
Therefore, $|t(k)|\to 1$ corresponds to a vanishing gap and maximal bandwidth, while $|t(k)|\to 0$ makes the bandwidth vanish entirely \cite{Hone1993}.
Furthermore, one may study the group velocity $v_\pm(\kappa,k)=\partial
\epsilon_\pm/\partial\kappa$, which measures the propagation speed of the
walker's wave packet \cite{Kittel2013}, whose maximum value is, precisely,
\begin{align}
  v_{\rm max}(k)= {}
  &
    \max_{\kappa}\left|v_{\pm}(\kappa,k)\right|
    \nonumber \\
    = {}
  &
    |t(k)|.
\end{align}
Thus, regardless of the quasimomentum, $|t(k)|$ sets a speed limit for the spreading of the quantum walker.

The quasienergy spectra as a function of $\kappa$ for the four point interactions are shown in Figure \ref{fig:fig5} for $k=1$.
In the case of $\delta$ [Fig. \ref{fig:fig5}(a)], we notice a gap between the two continuous bands.
The same gap is present in the  $\delta'$ spectra [Figure \ref{fig:fig5}(b)], except for a displacement in the quasienergy scale.
Conversely, for the $\chi$ interaction [Fig. \ref{fig:fig5}(c)], we obtain a gapless spectrum.
In the $\alpha$ interaction scenario [Fig. \ref{fig:fig5}(d)], we again obtain two separate bands, although with a smaller gap.
The quantities related to the spectra are shown for each point interaction considered in Fig. \ref{fig:fig6}.
In general, a large band gap is accompanied by slow-walker spreading in the lattice, alongside a small bandwidth, and vice versa.
Qualitatively, it is worth comparing the behavior of each frame with that in Fig. \ref{fig:fig2}. This lets us compare the spectral structure with the dynamical behavior of each point interaction.

The dependence of the quasienergy spectra on $k$ is seen in Figure \ref{fig:fig7}.
For small $k$, $\delta$ shows gapped flat bands [Figure \ref{fig:fig7}(a)], as $|t(k)| \to 0$ and $\Omega(\kappa,k)\to \pi/2$ independent of $\kappa$, so the bandwidth collapses and the gap saturates at its maximum value $\pi$.
Increasing $k$ leads to a gapless spectrum. 
In the $\delta'$ instance [Figure \ref{fig:fig7}(b)], we observe an opposite behavior, and $\chi$ shows [Fig. \ref{fig:fig7}(c)] a gapped spectrum at both ends of the $k$ axis and gapless at $k=1/\gamma$, where $|t|=1$. 
In contrast [Fig. \ref{fig:fig7}(d)], the $\alpha$ interaction shows a behavior similar to the $\delta$ case, but with a faster transition to a gapless spectrum.
The gaps observed in the spectra are analogous to those in solid-state physics, such as in the Kronig–Penney model \cite{Kittel2013}.

\section{Entanglement}
\label{sec:entanglement}

Entanglement is a quantum resource that arises from quantum correlations between different parts of a single quantum system associated with the non-separability of the quantum state \cite{Horodecki2009}.
It is necessary that the Hilbert space $\mathcal{H}$ of the total system can be decomposed into two or more parts, e.g., for an $N$-partite system $\mathcal{H}=\mathcal{H}_1\otimes\mathcal{H}_2\otimes...\otimes\mathcal{H}_N$.
In terms of DTQW models, as commented above, the Hilbert space of the CQW is written in terms of a tensor product, $\mathcal{H}_{\rm cqw}=\mathcal{H}_p \otimes \mathcal{H}_c$, and allows us to analyze the coin-position entanglement in this system  \cite{Carneiro2005}.
Thus, here we investigate the coin-position entanglement when using energy-dependent coins associated with point interactions.

\begin{figure*}
  \begin{center}
    \includegraphics[width=\textwidth]{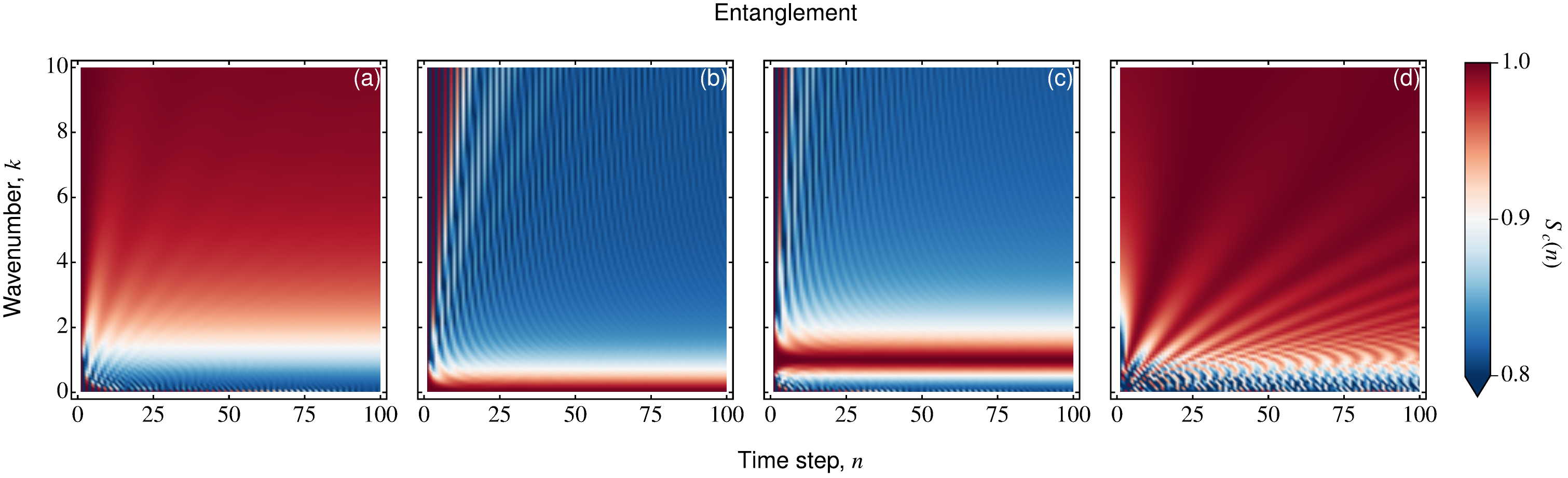}
  \end{center}
  \caption{
    Density plot of the time evolution of entanglement for $k\in[0,10]$,
    $\gamma=1$ and initial state
    $\ket{\phi(0)}=\ket{0}\otimes(\ket{-}+\ket{+})/\sqrt{2}$.
    We consider the cases: (a) $\delta$, (b) $\delta^{'}$, (c) $\chi$,
    and (d) $\alpha$.}
  \label{fig:fig8}
\end{figure*}

Since the state of the particle in CQW begins in a pure state and evolves under the action of a unitary operator, it remains pure after $n$ steps.
So, we can quantify the entanglement between the coin and position state
by the von Neumann entropy, which is defined by
\begin{equation}
  S_c(n)= -\sum_{i}\lambda_{i}\text{log}_{2}\lambda_{i},
\end{equation}
where $\lambda_{i}$ are the eigenvalues of the reduced density matrix of
the coin at time $n$.
The reduced density matrix of the coin is calculated using the partial trace operation on the position space, which gives
\begin{equation}
  \rho_{c}(n)=\Tr_{p}\left( |\phi(n)\rangle\langle\phi(n)| \right).
\end{equation}
In the present case of a unidimensional QW, $\rho_{c}(n)$ is associated with the $2\times 2$ matrix, i.e., has only two base states $\{\ket{+},\ket{-}\}$.
In Fig. \ref{fig:fig8}, we present the entanglement behavior for the $\delta$, $\delta^{'}$, $\chi$, and $\alpha$ interactions as a function of $k$.

For example, in the $\delta$ case, for $k=1$ (a balanced coin with $|r|=|t|=1/\sqrt{2}$, equivalent to the Hadamard walk), the entanglement approaches a limiting value $S_c \simeq 0.872$. We also observe that as $k$ increases, the oscillation of the entanglement values between the position and the coin state diminishes in such a way that it reaches a maximum at all times for large $k$'s.
On the other hand, in the $\delta^{'}$ case, for smaller $k$'s, the entanglement oscillates subtly.
However, for  $k>0$, the entanglement decreases as time passes.
As $k$ increases, the entanglement oscillations are greater at the outset of the dynamics and approach a limiting value for large times.
For the crossed case, we observe behavior similar to the $\delta^{'}$, with the entanglement being maximum for $k=1$. As $k$ increases, the entanglement oscillations also grow.
Lastly, the asymmetric case shows maximum entanglement for large $k$'s, and as $k$ becomes small, the entanglement exhibits oscillatory behavior.
Although the behavior differs for each point interaction, we see a pattern of entanglement approaching a limiting value. Of course, for each $k$ and each interaction, this limit assumes a different value.

Furthermore, considering a localized initial state, it is possible to obtain an asymptotic limit for entanglement \cite{Abal2006,Abal2006e}.
The time-evolved coin density matrix for the initial state $\rho_c(0)$ is, from Eq. \eqref{eq:kappa_matrix},
\begin{equation}
    \rho_{c}(n)=\int\frac{d\kappa}{2\pi}M^{n}(\kappa,k)\rho_{c}(0) \left[M^{\dagger}(\kappa,k)\right]^{n},
\end{equation}
where
\begin{equation}
   M(\kappa,k)= \left(\begin{array}{cc}
e^{i\kappa}t^{(+)}(k) & e^{i\kappa}r^{(-)}(k)\\
e^{-i\kappa}r^{(+)}(k) & e^{-i\kappa}t^{(-)}(k)
\end{array}\right).
\end{equation}
We have $\mathbf{M}(\kappa,k)=e^{-i{\phi(k)}/{2}} M(\kappa,k)\in SU(2)$, which allows the parametrization
\begin{equation}
\mathbf{M}(\kappa,k)=\cos\Omega(\kappa,k)\mathbbm{1}+i\sin\Omega(\kappa,k)\left(\vec{u}\cdot\vec{\sigma}\right)=e^{i\Omega(\kappa,k)\vec{u}\cdot\vec{\sigma}},
\end{equation}
with $\vec{\sigma}=\left(\sigma_x,\,\sigma_y,\,\sigma_z\right)$.
This allows for easier diagonalization in terms of the projectors
\begin{equation}
     P_{\pm} =\frac{1}{2}\left(\mathbbm{1}\pm\vec{u}\cdot\vec{\sigma}\right).
\end{equation}
For notational simplicity, we omit the quasimomentum and wavenumber dependence on both $\vec{u}$ and $P_\pm$.
Thus, the time-evolution of $\rho_{c}(n)$ is given by
\begin{align}
  \label{eq:int}
  \rho_{c}(n)  = {}
  &
    \int \frac{d\kappa}{2\pi}\left[P_{+}\rho_{c}(0)P_{+}
    +e^{2i\Omega(\kappa,k)n}P_{+}\rho_{c}(0)P_{-}
    \right. \nonumber\\
  &
    \left.+e^{-2i\Omega(\kappa,k)n}P_{-}\rho_{c}(0)P_{+}
    +P_{-}\rho_{c}(0)P_{-}\right].
\end{align}

For $0<|t(k)|\leq 1$, $\Omega(\kappa,k)$ is a non-constant function of $\kappa$, and the crossed terms vanish in the long-time limit \cite{Nayak2000}. 
In the limit $|t(k)|=0$, however, $\Omega(\kappa,k)=\pi/2$ is independent of $\kappa$.
In this scenario, $S_c(n)$ oscillates without converging and the asymptotic values are therefore limits that are approached ever more slowly as $|t(k)|\to0$.
This is the origin of the persistent fringes visible at small $k$ for $\delta$ and $\alpha$, and at large $k$ for $\delta'$ and $\chi$ in Fig. \ref{fig:fig8}.

\begin{figure}[b]
  \centering
  \includegraphics[width=\columnwidth]{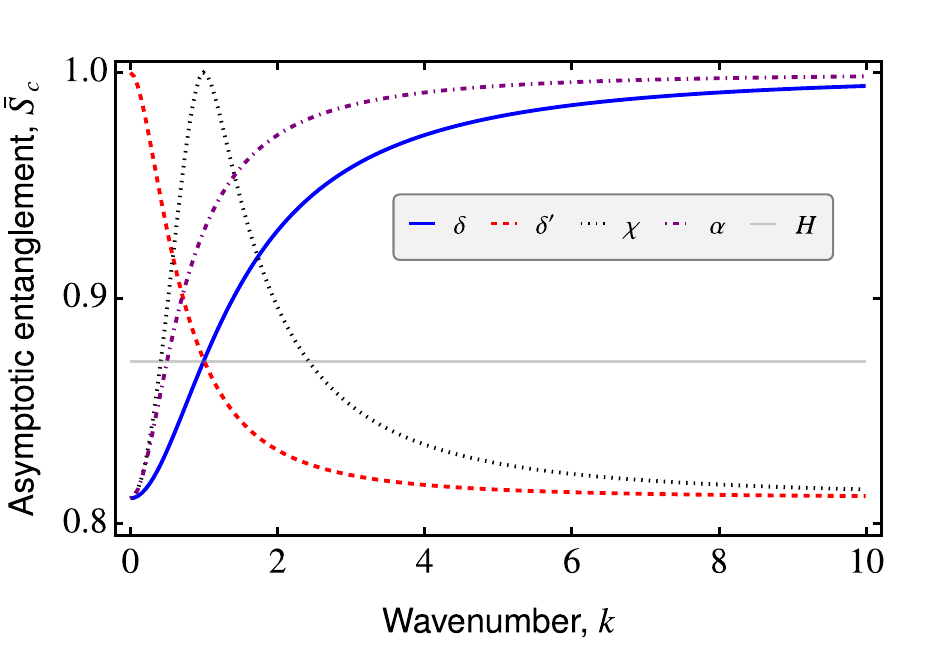}
  \caption{
   Asymptotic entanglement for the point interactions with $\gamma=1$ and initial state $\ket{\phi(0)}=\ket{0}\otimes\left(\ket{-}+\ket{+}\right)/\sqrt{2}$: $\delta$ (solid blue line), $\delta'$ (dashed red line), $\chi$ (dotted black line), and  $\alpha$ (dot-dashed purple line). The faded gray line represents the known asymptotic limit for the Hadamard walk.
    }
  \label{fig:fig9}
\end{figure}

Considering the initial state, $\ket{\phi(0)}=\ket{0}\otimes(\ket{-}+\ket{+})/\sqrt{2}$, diagonalization followed by integration leads to the eigenvalues \cite{Orthey2019,Ide2011}
\begin{equation}
\label{eq:lambdapm}
    \bar{\lambda}_{\pm}
    =\lim_{n \to\infty}\lambda_{\pm}
    =\frac{1}{2}
    \left(1\pm\frac{|r(k)|}{1+|r(k)|}\right),
\end{equation}
with $|r(k)|=\sqrt{1-|t(k)|^2}$.
Considering the interval $\left|t(k)\right|<1$, we obtain from Eq. \eqref{eq:lambdapm} asymptotic entanglements in the interval of $\bar{S}_c\in\left[2-3/4\log_23,1\right]\approx\left[0.811,1\right]$.
For $\left|t(k)\right|=1/\sqrt{2}$, we obtain the known value of $\bar{S}_c \approx 0.872$ for the Hadamard walk.
Entanglement for these long-time values is found in Fig. \ref{fig:fig9} for the four point interactions considered.
Comparing with Fig. \ref{fig:fig2}, we find a monotonic relationship between the transmission probability and the asymptotic entanglement.
This explains why the long-time values in the four panels of Fig. \ref{fig:fig8} differ so little and shows that the coin–position entanglement is a considerably weaker probe of the underlying point interaction than the transmission coefficient itself.

Finally, the manner in which the asymptotic value is approached carries spectral information of its own.
The crossed terms in Eq. \eqref{eq:int} oscillate rapidly, so the only surviving contribution comes from stationary points satisfying $\partial \Omega(\kappa,k)/\partial\kappa = 0$.
These stationary points occur at $\kappa_1 = -\arg[\omega]$ and $\kappa_2 = \pi - \arg[\omega]$, yielding $\Omega(\kappa_1, k) = \arccos[|t(k)|]$ and $\Omega(\kappa_2, k) = \pi - \arccos[|t(k)|]$.
For the initial state used here, the residual contribution decays as $n^{-1/2}$ for $\alpha$ and as $n^{-3/2}$ for $\delta$, $\delta'$, and $\chi$, and oscillates with frequency $W(k)$.
Thus, the transient oscillations of the coin-position entanglement provide a dynamical probe of the quasienergy gap.

\section{Conclusions}
\label{sec:conclusion}

We have introduced DTQW, whose coin operator is inherited from the scattering matrix of an energy-dependent point interaction, $\mathbf{C}_c(k) =  \sigma_x S(k)$, making the walk a discrete-time realization of a quantum scattering problem with a periodic configuration of potentials. Working in the SQW picture, the natural setting for energy-dependent scattering amplitudes, and analyzing entanglement in the unitarily equivalent CQW picture, which carries the bipartite structure, allowed us to retain the physical transparency of the former and the computational convenience of the latter. In this context, the walker's wavenumber is a control parameter with clear physical significance for the coin operation, enabling extensive investigation in parameter space.

This equivalence lets us examine various figures of merit, namely spectra, entanglement, and spreading—from a scattering perspective. The transmission modulus determines the minimum quasienergy gap, bandwidth, and maximum group velocity exactly. For the separable initial state considered, the asymptotic coin-position entanglement is likewise determined by the reflection modulus and, hence, the transmission modulus. The quasienergy spectrum consists of two bands whose minimum gap, bandwidth, and maximum group velocity are all determined by the transmission modulus. Each dependency manifests differently across point interactions, yielding distinct profiles that span the full range of DTQW behaviors. We further analyze the walker's spatial spreading, which is directly linked to the energy spectra: the minimum band gap and the transmission coefficient are directly related.
Consequently, increasing transmission simultaneously reduces the minimum gap, increases the bandwidth, and enhances the maximum propagation velocity. 
We also study coin-position entanglement, enabling us to probe the entanglement between the two degrees of freedom of the SQW, an interplay that exhibits distinct signatures for each interaction. Finally, we examine the asymptotic limit of this entanglement, revealing how the wavenumber governs its long-time behavior for all four potentials. The scattering coefficients likewise fix this quantity.

Experimentally, the model requires beam splitters with wavelength-dependent reflectivity or a coin whose splitting ratio and relative phase depend on the walker's wavelength. Discrete-time photonic walks with fully tunable coins are now routine, both in time-multiplexed fiber loops \cite{Schreiber2010,Broome2010} and in femtosecond waveguide lattices, where the splitting ratio of a directional coupler is set by the evanescent overlap of the two guided modes and is therefore intrinsically dispersive \cite{Szameit2007,Peruzzo2010,Meany2015}.
The signature we would single out for measurement is the locking between the entanglement oscillation frequency and the band gap, both tunable through a single input wavelength.

\section*{Acknowledgements}
We thank Renato Moreira Angelo for interesting discussions.
This work was partially supported by Coordenação de Aperfeiçoamento de
Pessoal de Nível Superior (CAPES, Finance Code 001).
It was also supported by Conselho Nacional de Desenvolvimento Científico
Tecnológico (CNPq).
F.M.A. acknowledges financial support from Fundação Araucária Project No. 305 and CNPq Grant No. 313124/2023-0.

\appendix

\section{Flux conservation in a point interaction}
\label{sec:appendix}

In standard quantum mechanics, the probability current is defined as
\cite{Book.1994.Shankar}
\begin{equation}
  J(x)=\frac{1}{2i}
  \left[
    \psi^{*}(x)\psi'(x)-\psi(x)\psi'^{*}(x)
  \right],
  \label{eq:fluxo}
\end{equation}
which can be written as
\begin{equation}
  J(x)=\frac{1}{2i}\Psi^{\dagger}(x)\mathbb{S}\Psi(x),
  \label{eq:symplectic}
\end{equation}
with $\Psi(x)$ given by \eqref{eq:Psi} and
\begin{equation}
  \mathbb{S}=
    \begin{pmatrix*}[r]
      0 & 1\\
      -1 & 0
    \end{pmatrix*},
\end{equation}
being the canonical symplectic matrix.
In this representation, the probability current is written as a sesquilinear form, and current conservation follows from the invariance of this form under the boundary transformation $\Gamma$.
Using Eq. \eqref{eq:bc} for a point interaction located at $x=0$, we have the following:
\begin{align}
  J(0^{+})
  = {}
  &
    \frac{1}{2i}
    \Psi^{\dagger}(0^{+}) \mathbb{S} \Psi(0^{+}),
    \nonumber \\
    = {}
  &
    \frac{1}{2i}
    \Psi^{\dagger}(0^{-})  \Gamma^{\dagger} \mathbb{S} \Gamma \Psi(0^{-}).
\end{align}
Now, using  $|\omega|=1$ and $a d - b c =1$, we have
$
  \Gamma^{\dagger} \mathbb{S} \Gamma = \mathbb{S}.
$
Therefore,
\begin{align}
  J(0^{+})
  ={}&
       \frac{1}{2i}
       \Psi^{\dagger}(0^{-}) \mathbb{S} \Psi(0^{-})
       \nonumber\\
  ={}&
       J(0^{-}).
\end{align}
Hence, if Eq. \eqref{eq:bc} is satisfied with
$a,b,c,d\in\mathbb{R}$ and $|\omega|=1$, the probability current is
conserved across the point interaction.

%

\end{document}